\documentstyle[12pt,emlines]{article}
\begin{document}

\begin{center}
\large \bf {Superconductivity in two-band non-adiabatic \\
systems.}
\end{center}

\begin{center}
\large \bf {M.E. Palistrant and V.A. Ursu }
\end{center}

Theory of superconductivity in two-band non-adiabatic systems with strong
electron correlations in the linear approximation over non-adiabaticity  is
built in the article. Having assumed weak electron-phonon interaction analytical
expressions for the vertex functions $P_{Vn}$and "intersecting " functions
$P_{Cn}$ for each energy band ($n = 1, 2$) are obtained. Mass operators of Green
functions with account of diagrams with intersection of two lines of electron-
phonon interaction (exceeding the frameworks of Migdal theorem) are derived as
well as main equations of theory of superconductivity in two-band systems.
Analytical formulas for the temperature of superconducting transition $T_C$  are
obtained as well as renormalization of the quantities which enter these formulas
because of  effects of non-adiabaticity and strong electron correlations.
Dependence of $T_C$ and coefficient of isotope effect $\alpha$ on Migdal
parameter $m = \omega_0/\varepsilon_F$ is studied. Overlapping of energy bands
on Fermi surface with effects of non-adiabaticity at low values of transferred
momentum $(q \ll 2p_F)$ is shown to obtain high values of  $T_c$ even assuming
weak electron-phonon interaction.\\

\begin{center}
{\bf 1. Introduction}
\end{center}

There have been 20 years since the discovery of high -$T_C$ materials
\cite{Berdnorz}. There have been made thorough experimental and theoretical
studies of properties of these materials in normal and superconducting state. It
is established the significant distinction of physical properties in comparison
with the case of ordinary superconductors.  New high-$T_C$ materials have some
peculiarities such as strong anisotropy of crystalline properties (layered
structure), strong electron correlations, strong electron-phonon interaction,
presence of the points with high symmetry in the momentum space, overlapping of
energy bands on Fermi surface, small values of Fermi energy.  Sophisticated
nature of these systems makes it difficult to solve one the main problems of
modern physics - to find the real mechanism of high-$T_C$ superconductivity.
Thorough consideration of all peculiarities in these materials is one of the
possible solutions of this problem as well as study of their influence on
thermodynamic and magnetic properties of system.

In order to study effect of overlapping of energy bands on Fermi surface, for
example, which results in two energy gaps there was used the model proposed by
Moscalenco \cite{Moskalenko} (see also Suhl et all \cite{Suhl}). Theory of
superconductivity for ordinary superconductors (pure and doped) which is based
on this model was build long before the discovery  of high-$T_C$ materials (see
history of this development in  \cite{Palistrant}, \cite{Kon}).

The application of this theory to the high-$T_C$ materials was done by us and
other authors (see references in \cite{Moskalenko_1}). A great number of
experimental data over superconducting properties of the metal-oxide compounds
are snowed to be understood in the frameworks of this theory
\cite{Moskalenko_1}.

From the other hand, there are some questions and the role of non-adiabaticity
should be explained as well as of strong electron correlations in high-$T_C$
materials. More specifically, we have the case with low values of Fermi energy
$(\varepsilon_F \ll \omega_D)$ or when it is compared with Debye energy
$(\varepsilon_F \sim \omega_D)$, and great values of Coulomb interaction
between electrons.  These factors can not  favor the high-$T_C$
superconductivity within the frameworks of the BCS-Bogolyubov-Eliashberg-Migdal
theory. In addition, this theory is not valid to describe non-adiabatic systems
when Migdal theorem is broken down \cite{Migdal}. That is why there is
necessity to take into consideration vertex and "intersecting" diagrams over
electron-phonon interaction (additional multi-particle effects) at determining
Green functions.

There is great number of works which study superconductivity in non-adiabatic
systems and build the theory of superconductivity by going beyond Migdal theorem
(see review  \cite{Danilenko}). The vertex functions at $(\varepsilon_F \sim
\omega_D)$ is followed from these studies to be negative and as result to
decrease the coupling of electron-phonon interaction which in its turn does not
give the high values of the temperature of superconducting transition $T_C$.
The opposite picture we have when the transferred momentum of electron-phonon
interaction is small $(q \ll 2p_F)$.  Such a specific case arises when strong
electron correlation are taken into account \cite{Kulic}, \cite{Zeyher}. A
non-adiabatic system with half-filling of energy band have been studied in
works \cite{Pietronero}, \cite{Grimaldi} where the cutoff of the
electron-phonon interaction over the transferred momentum $q_c \ll 2p_F$. In
this way the vertex $P_V$ and intersecting $P_C$  functions are positive and
favor the significant increase of the coupling of electron-phonon interaction,
rising as a result the temperature of superconducting transition up to the
values of high-$T_C$ materials. Positive values of $P_V$ and $P_C$ are also
possible in the non-adiabatic systems with van Hove singularities in the
electron energy spectrum. This is observed clearly, for instance, in the case of
the "extended" singularity in electron energy spectrum \cite{Palistrant_1},
which is found experimentally in yttrium ceramics \cite{Abrikosov}.

Note that studies of thermodynamic properties of one-band superconducting non-
adiabatic system with variable density of charge carriers (pure and with
magnetic impurity) have been performed by one of the authors
\cite{Palistrant_2}, \cite{Palistrant_3}. These studies show that non-
adiabaticity favors the superconductivity of one-band systems as well as strong
electron correlations.

Intensive investigations of new superconducting materials have been made for
last five years. Compound $MgB_2$ with $T_c \sim 40 k$ \cite{Nagamatsu} is of
great interest. This compound has some anomalies amongst its physical properties
(see, for example, reviews \cite{Confield}, \cite{Bouquet}). These anomalies are
established to be caused by properties of system anisotropy and, in particular,
by overlapping of energy bands on Fermi surface. The theory of superconductivity
with energy bands that overlap on Fermi surface\cite{Moskalenko},
\cite{Confield} - \cite{Dolgov} has been put into the kernel of description of
the properties of $MgB_2$ compound. The effects of non-adiabaticity can play
significant role in establishing properties of $MgB_2$ compound because Migdal
parameter in these compounds $m = \frac{\omega_D}{\varepsilon_F} \approx 0.1$
\cite{Aleksandrov}.

Therefore, modern high-$T_C$ superconductors (oxide metals, fullerens, organic
systems, boride magnesium) are non-adiabatic systems.  In addition to this
electron energy spectrum has essential peculiarity - overlapping of
energy bands on Fermi surface what is proved by band calculations (see, for
example, 24 -27).  All this encourages us to build theory of superconductivity
in two-band non- adiabatic systems. This article is aimed to perform this task.
The article is designed in the following way:\\ Having based on the Hamiltonian
of Frolich type the equations for temperature electron Green functions are
obtained in section 2 as well as equations for mass operators $M$
and $\Sigma$. Transition to the band representation has been performed as well
$n \vec k \Omega$ - representation.\\ Definitions of vertex $P_{Vn}$ and
"intersecting" $P_{Cn}$ functions of each energy band are given in section 3.
The system of equation for order parameters of two-band system is also derived
here.\\ The equation for the temperature of superconducting temperature $T_C$
has been obtained in section 4 and its solution have been found as well. In
addition the coefficient of isotope effect $\alpha$ is determined also.\\
Numerical calculations are performed in section 5 as well as the analysis of
results.\\

\begin{center}
{\bf 2. Hamiltonian of the system and Green functions.}
\end{center}

We start with Frolich type Hamiltonian
\begin{equation}
H = H_0 + H_1, %1
\end{equation}
where $H_0$ is Hamiltonian of non-interacting electrons and phonons, and $H_1$
is electron-phonon interaction determined by expression
\begin{equation}
H_1 = \sum_{\sigma} \int d \vec x \psi_{\sigma}^{+} (\vec x) \psi_{\sigma}
(\vec x) \varphi (\vec x).%2
\end{equation}

Here $\psi_{\sigma}^{+} (\vec x)$ and $\psi_{\sigma} (\vec x)$ - are creation
and annihilation operators of electrons with spin $\sigma$ at the point $\vec
x$.  Having based on the Hamiltonian (1) and using diagrammatic method of the
perturbation theory \cite{Abrikosov_1} we obtain the system of equations for normal
$G(x, x')$ and abnormal $\tilde F(x, x')$ Green function:
$$ G (x, x') = G^0 (x, x') + \int \int
dx_1 dx_2 G^0 (x, x_1) M (x_1, x_2) G (x_2, x') -
$$
$$ -\int \int dx_1 dx_2 G^0
(x, x_1) \Sigma (x_1, x_2) \tilde F (x_2, x'),
$$
$$ \tilde F (x, x') = \int dx_1
dx_2 G_0 (x_1, x) \tilde \Sigma(x_1, x_2) G (x_2, x') +
$$
\begin{equation}
+ \int dx_1 dx_2 G^0 (x_1, x) M (x_2, x_1) \tilde F (x_2, x').%3
\end{equation}

Because of absence of spin interactions in (1) to obtain  the system (3) we used
the following relationships:
\begin{equation}
G_{\alpha \beta} (x, x') = \delta_{\alpha, \beta} G (x, x');\quad
\tilde F_{\alpha \alpha'} (x, x') = - g_{\alpha, \alpha'} \tilde F (x, x'),%4
\end{equation}
where
\begin{equation}
g_{\alpha, \alpha'} = \delta_{\alpha, - \alpha'} (\delta_{\alpha \uparrow} -
\delta_{\alpha \downarrow}) = - g_{\alpha', \alpha}.%5
\end{equation}
Mass operators diagonal $M$ and non-diagonal $\Sigma$
at temperatures close to the critical $(T \sim T_C)$ have the
following form:
$$
M(x_1, x_2) = - D (x_1, x_2) G (x_1, x_2) + \int \int d x_3 d x_4 D (x_4, x_2)
D(x_1, x_3)\times
$$
\begin{equation}
\times G (x_1, x_4) G (x_4, x_3) G (x_3, x_2),
\end{equation}%6
$$
\Sigma (x_1, x_2) =  - D (x_1, x_2) F (x_1, x_2) + \int \int d x_3 d x_4 D
(x_1, x_3) D(x_4, x_2)\times
$$
$$
\times \left[G (x_1, x_4) G (x_2, x_3) F (x_4, x_3) + G (x_3, x_4) F(x_1, x_4)
G (x_2, x_3) + \right.
$$
\begin{equation}
+\left. F (x_3, x_2) G (x_1, x_4 ) G (x_4, x_3) - F(x_1, x_3) \tilde F
(x_3, x_4) F(x_4, x_2)\right] .
\end{equation}%7
As follows from (6) and (7) there are diagrams with intersection of two lines
of electron -phonon interaction as well as diagrams that enter the mass
operators in adiabatic systems. Account of these terms represents the vertex
corrections and exceedance beyond bounds of Migdal theorem \cite{Migdal}.  We
go in (3), (6) and (7) to  $n \vec k \Omega$ representation in accordance with
decompositions of Green functions:
$$
G(x, x') = \frac{1}{\beta} \sum_{\vec k \vec k'} \sum_{nm} \sum_{\Omega} G_{nm}
(\vec k, \vec k', \Omega) \psi_{n \vec k} (\vec x) \psi_{m \vec k'}^{*} (\vec
x') e^{- i \Omega (\tau - \tau')},
$$
$$
F(x, x') = \frac{1}{\beta} \sum_{\vec k
\vec k'} \sum_{nm} \sum_{\Omega} F_{nm} (\vec k, \vec k',\Omega) \psi_{n \vec
k} (\vec x) \psi_{m \vec k'} ( \vec x') e^{- i \Omega (\tau - \tau')},
$$
\begin{equation}
\tilde F(x, x') = \frac{1}{\beta} \sum_{\vec k \vec k'} \sum_{nm} \sum_{\Omega}
\tilde F_{nm} (\vec k, \vec k',\Omega) \psi_{\vec n \vec k}^{*} (\vec x)
\psi_{m' \vec k'}^{*} (\vec x') e^{- i \Omega (\tau - \tau')},%8
\end{equation}
and for mass operators as well
$$
M(x, x') = \frac{1}{\beta} \sum_{\vec k \vec k'} \sum_{nm} \sum_{\Omega} M_{nm}
(\vec k, \vec k',\Omega) \psi_{n \vec k} (\vec x) \psi_{m \vec k'}^{*} (\vec x')
e^{- i \Omega (\tau - \tau')},
$$
$$
\Sigma (x, x') = \frac{1}{\beta} \sum_{\vec k \vec k'} \sum_{nm} \sum_{\Omega}
\Sigma_{nm} (\vec k, \vec k',\Omega) \psi_{n \vec k} (\vec x) \psi_{m \vec k'}
(\vec x') e^{- i \Omega (\tau - \tau')},
$$
\begin{equation}
\tilde \Sigma(x, x') = \frac{1}{\beta} \sum_{\vec k \vec k'} \sum_{nm}
\sum_{\Omega} \tilde \Sigma_{nm} (\vec k, \vec k',\Omega) \psi_{n \vec k}^{*}
(\vec x) \psi_{m' \vec k'}^{*} (\vec x') e^{- i \Omega (\tau - \tau')},%9
\end{equation}
where $\psi_{n \vec k} (\vec x) = e^{i \vec k \vec x} U_{n \vec k}(\vec x)$ is
Bloch function.

Having substituted (8) and (9) in (3) and performed all necessary actions we
obtain the system of equations for $G_{nm} (\vec k \vec k', \Omega)$ and
$\tilde F_{nm}(\vec k \vec k', \Omega)$. Solution of this system of equations
and thermodynamic properties of two-band superconductors without effects of
non-adiabaticity are presented in works \cite{Kochorbe}, \cite{Palistrant_4}.
In this way all possible superconducting pairings between electrons in each
energy band and between electrons from different bands are taken into
consideration.

Hereinafter we restrict ourselves by approximation diagonal indexes over bands:
\begin{equation}
G_{nm}(\vec k, \vec k',\Omega) = \delta_{\vec k \vec k'} \delta_{nm} G_n
(\vec k, \Omega);\quad \tilde F_{nm} (\vec k, \vec k',\Omega) = \delta_{\vec k,
- \vec k'} \delta_{nm} \tilde F_n (\vec k, \Omega).%10
\end{equation}

If we do not take into consideration the effects, connected with non-
adiabaticity of a system this approximation leads to two-band model of
Moscalenco \cite{Moskalenko}. This model correponds to the formation of the
cooper pairs inside each energy band and their tunneling from one energy band
to another one. In approximation (10) solution of equation(3) can be presented
in the form:
\begin{equation}
G_n (\vec k, \Omega) = -\frac{i Z_n (\Omega) \Omega + \tilde \varepsilon_n
(\vec k)}{A_n (\vec k, \Omega)},\quad \tilde F_n (\vec k, \Omega) =
\frac{\tilde \Sigma_n (k, \Omega)}{A_n (\vec k, \Omega)},\quad F_n (\vec k,
\Omega) = \frac{\Sigma_n (\vec k, \Omega)}{A_n (\vec k, \Omega)},%11
\end{equation}
where
$$
A_n (\vec k, \Omega) = \Omega^2 Z_n^2 (\Omega) + \tilde
\varepsilon_n^2 (\vec k) + \Sigma_n (\vec k, \Omega) \tilde \Sigma_n (\vec k,
\Omega),
$$
\begin{equation}
Z_n (\Omega) = 1 - \frac{1}{\Omega} Im M_n (\vec
k, \Omega),\quad \tilde \varepsilon_n (\vec k) = \varepsilon_n (\vec k)+ Re M_n
(\vec k, \Omega).%12
\end{equation}

$M_n$ and $\Sigma_n$ are mass operators in $n \vec k \Omega$ representation that
are determined by expressions (6) and (7) and  take into account diagrams which
breaks down Migdal theorem. \\

\begin{center}
{\bf 3. Mass operators and vertex functions.}
\end{center}

Having gone to $n \vec k \Omega$ - representation in definitions of mass
operators (6) and (7) in accordance with the decompositions  (8) and (9) we
obtain:
$$
M_m (\vec p, \Omega) = - \frac{1}{\beta} \sum_{\vec p_1} \sum_{\Omega_1}
\sum_n D(\vec p - \vec p_1, \Omega - \Omega_1) g_{mn}^{2} (p, p_1) G_n (\vec
p_1, \Omega_1) +
$$
$$
+ \frac{1}{\beta^2} \sum_{\vec p_1\, p_2} \sum_{\Omega_1\,\Omega_2}
D (\vec p - \vec p_1, \Omega - \Omega_1) D(\vec p - \vec p_2, \Omega -
\Omega_2) g_{mm}^{2}(\vec p, \vec p_1) g_{mm}^{2}(\vec p, \vec p_2).
$$
\begin{equation}
G_m (\vec p_1, \Omega_1) G_m (\vec p_1 + \vec p_2 - \vec p, \Omega_1 + \Omega_2
- \Omega) G_m (p_2, \Omega_2),%13
\end{equation}
$$
\Sigma_m (\vec p \Omega) =
-\frac{1}{\beta} \sum_{\vec p_1} \sum_{\Omega_1} \sum_{n_1} D(\vec p - \vec
p_1, \Omega - \Omega_1) g_{mn_1}^{2} (\vec p, \vec p_1) F_{n_1} (\vec p_1,
\Omega_1) +
$$
$$
+ \frac{1}{\beta^2}
\sum_{\vec p_1, \vec p_2} \sum_{\Omega_1,\Omega_2} D(\vec p - \vec p_1, \Omega
- \Omega_1) D(p - p_2, \Omega - \Omega_2) g_{mm}^{2} (\vec p, \vec p_2)
g_{mm}^{2} (\vec p, p_1) \left[G_{m} (\vec p_2, \Omega_2)\right. \times
$$
$$
\times F_{m}(\vec p_1, \Omega_1) G_{m} (\vec p_2 + \vec p_1 - \vec p, \Omega_2
+ \Omega_1 - \Omega) +
$$
$$
+ F_{m} (\vec p_1, \Omega_1) G_{m} (\vec p_2,
\Omega_2) G_{m}(\vec p_2 - \vec p_1 - \vec p, \Omega_2 - \Omega_1 - \Omega) +
$$
\begin{equation}
\left.+ F_{m} (\vec p_1, \Omega_1) G_{m} (\vec p - \vec p_1 - \vec p_2,
\Omega - \Omega_1 - \Omega_2) G_{m} (- \vec p_2,  - \Omega_2)\right].%14
\end{equation}

Here there was used the approximation over band indexes (10) and approximation
for matrix elements of electron-phonon  interaction that corresponds to the
tunneling of electron pairs as a whole from one energy band to another (
Moscalenco model \cite{Moskalenko}):

\begin{equation}
g_{mn}^{2} (\vec p, \vec p_1) = g^2(\vec p,\vec p_1) \left|\chi (m  \vec p_F,\,n
\vec p_{1 F})\right|^2 , %15
\end{equation}
where
\begin{equation}
\chi (m \vec p_F, n \vec p_{1 F}) = \int \limits_{V_0} d \vec r  U_{m \vec
p_F}^{*} (\vec r) U_{n \vec p_{1 F}} (\vec r)%16
\end{equation}
$m; n = 1, 2$, $V_0$ is volume of the basic cell.

Assume that strong electron correlations \cite{Kulic},\cite{Zeyher} can lead to
low values of transferred momentum $\vec q$ in electron-phonon interaction. In
accordance with this assumption we represent $g_{mn}^{2}(\vec p \vec p\,')$ in
the following form:
\begin{equation}
g_{mn}^{2} (\vec p, \vec p\,') = g_{mn}^{2} \eta \theta (q_c - |\vec p -
\vec p\,'|),%17
\end{equation}
where $q_c$ is cut-off momentum of electron-phonon interaction, $\eta$ is
determined by relation
\begin{equation}
\eta \ll \theta (q_c - |\vec p - \vec p '|) \gg_{FS} = 1,%18
\end{equation}
where $\ll g_{mn}^{2} (p p')\gg_{FS} = g_{mn}^{2}$.
Here $\ll .... \gg_{FS}$ means averaging over Fermi surface.

We choose phonon Green function $D (\Omega \Omega_1)$ with simple Einstein
spectrum (characteristic frequency of which is $\omega_0$):
\begin{equation}
D(\Omega, \Omega_1) = - \frac{\omega_0^2}{(\Omega - \Omega_1)^2 +
\omega_0^2},%19
\end{equation}
and introduce  vertex functions $(n = 1,2)$:
$$
P_{Vn} (\vec p, \vec p_1, \Omega, \Omega_1) = - \frac{1}{\beta} \frac{\eta}{N_n}
\sum_{\vec p_2} \sum_{\Omega_2} \theta (q_c - |\vec p - \vec p_2|) D (\Omega,
\Omega_2) G_n(\vec p_2, \Omega_2) \times
$$
$$
\times G_n (\vec p_1 + \vec p_2 - \vec p, \Omega_1 +
\Omega_2 - \Omega),
$$
$$
P_{Cn} (\vec p, \vec p_1, \Omega, \Omega_1) = - \frac{1}{\beta} \frac{\eta}{N_n}
\sum_{\vec p_2} \sum_{\Omega_2} \theta (q_c - |\vec p - \vec p_2|) D (\Omega,
\Omega_2) G_n(\vec p_2, \Omega_2) \times
$$
$$
\times G_n (\vec p_1 - \vec p_2 - \vec p, \Omega_1 -
\Omega_2 - \Omega),
$$
$$
P_{Vn}^{'} (\vec p, \vec p_1, \Omega, \Omega_1) = - \frac{1}{\beta}
\frac{\eta}{N_n}\times
$$
\begin{equation}
\times \sum_{\vec p_2} \sum_{\Omega_2} \theta (q_c - |\vec p - \vec p_2|) D
(\Omega, \Omega_2) G_n (\vec p - \vec p_1 - \vec p_2, \Omega - \Omega_1 -
\Omega_2) G_n (- \vec p_2, -\Omega_2).%20
\end{equation}
Having used further the representation (17), (20)  and performed averaging over
Fermi surface in equation (13) and (14), we derive
$$
\ll M_m (\vec p \Omega)\gg_{FS} = M_m (\Omega) =-
$$
$$
- \frac{1}{\beta} \sum_{\vec p_1} \sum_{\Omega_1}  g_{mm}^{2} D (\Omega,
\Omega_1) G_m (\vec p_1, \Omega_1) \left [1 + g_{mm}^{2} P_{Vm} (Q_c, \Omega,
\Omega_1)\right] -
$$
\begin{equation}
- \frac{1}{\beta} \sum_{\vec p_1, \Omega_1} g_{mn}^{2} D(\Omega, \Omega_1)
G_n(\vec p_1, \Omega_1),
%21
\end{equation}
$$
\Sigma_1 (\Omega) = -\frac{1}{\beta} \sum_{\vec p_1} \sum_{\Omega_1} D(\Omega,
\Omega_1) \widetilde V_{11} (\Omega, \Omega_1) F_1 (\vec p_1, \Omega_1) -
\frac{1}{\beta} \sum_{\vec p_1, \Omega_1}D (\Omega, \Omega_1) g_{12}^{2} F_2
(\vec p_1, \Omega_1),
$$
$$
\Sigma_2 (\Omega) = - \frac{1}{\beta} \sum_{\vec p_1} \sum_{\Omega_1} D
(\Omega, \Omega_1) g_{21}^{2} F_2 (\vec p_1, \Omega_1) -
$$
\begin{equation}
- \frac{1}{\beta} \sum_{\vec p_1} \sum_{\Omega_1} D (\Omega, \Omega_1) \tilde
V_{22} (\Omega, \Omega_1) F_2 (\vec p_1, \Omega_1),%22
\end{equation}
where
\begin{equation}
\widetilde V_{11} (\Omega, \Omega_1) = g_{11}^{2} + N_1 g_{11}^{4} \left[
P_{V_1} (Q_c, \Omega, \Omega_1) + P_{V_1}^{'} (Q_c, \Omega, \Omega_1) + P_{c_1}
(Q_c, \Omega, \Omega_1)\right]
\end{equation}%23
As it follows from (21) - (23) the vertex and "cross - intersected" functions
contribute to the intraband electron - phonon interaction constants $g_{11}^{2}$
and $g_{22}^{2}$ without influencing the interband interaction $g_{12}^{2}$.
The value $\tilde V_{22}$ is obtained replacing $1 \rightarrow 2$ in the
formula (23).

Expressions $P_{V, Cn} (Q_c, \Omega, \Omega_1)$ correspond to formulas averaged
over Fermi surface, $Q_c = q_c / 2p_F$.

Hereinafter, in performing calculations we use simple dispersion law of electron
energy for $n$ - th band
\begin{equation}
\varepsilon_n (\vec p) = \zeta_n + \frac{p^2}{2_{m_n}} - \mu. % (24)
\end{equation}
Let's go in definitions of function $P_{V cn}$ (20)  from summation over $\vec
p_2$ and $\Omega_2$ to integration over energy and frequency in the standard way
(as it was done at $T = 0$):
$$
\frac{1}{\beta} \sum_{\vec p_2} \sum_{\Omega_2} \Phi_n (\vec p_2, \Omega_2) =
N_n \int \limits_{0}^{2 \pi} \frac{d \varphi}{4 \pi} \int \limits_{0}^{\pi} sin
\theta d \theta \int \limits_{-\mu_n}^{W_n- \mu_n} d \varepsilon_{p_2}
\frac{1}{2 \pi} \int \limits_{- \infty}^{\infty} d \Omega_2 \Phi_n(\vec p_2,
\Omega_2)
$$
where
\begin{equation}
N_n = \frac{m_n \vec p_{F_n}}{2 \pi^2},\,\,\,\mu_n = \mu - \zeta_n.%25
\end{equation}
Integration over $\Omega_2$ with infinite limits is caused by  approximation of
weak coupling of electron-phonon interaction $(\omega_0/T_c \gg 1)$. Integration
over angular variables  is performed in accordance with lowness of transferred
momentum $q (q \ll 2p_F)$. Here we have used the method of calculation that was
developed for this case in works \cite{Grimaldi}, \cite{Palistrant_2},
\cite{Palistrant_3}. In this way we take into consideration for three-
dimensional system at $\theta_c \ll \omega_0/\mu_n$ expressions
$$
P_{Vn} (Q_c, \Omega, \Omega_1) = \frac{\omega_0 A_n (\Omega, \Omega_1)}{(\Omega
- \Omega_1)^2} - \frac{\omega_0 E^2}{(\Omega - \Omega_1)^4} \left[ A_n (\Omega,
\Omega_1) - (\Omega - \Omega_1)^2 B_n (\Omega, \Omega_1)\right] \frac{1}{2}
Q_c^4,
$$
$$
P_{Cn} (Q_c, \Omega, \Omega_1) = \frac{\omega_0 A_n (\Omega, - \Omega_1)}{
(\Omega + \Omega_1)^2} - \frac{\omega_0 E^2}{(\Omega + \Omega_1)^4} \left[
A_n (\Omega, - \Omega_1) - (\Omega + \Omega_1)^2 B_n (\Omega, - \Omega_1)
\right]\times
$$
\begin{equation}
\times\frac{22}{3} Q_c^4 + \frac{\omega_0 E}{(\Omega + \Omega_1)^2} C_n
(\Omega, - \Omega_1) Q_c^2,%26
\end{equation}
where $E = 2 \varepsilon_F$,
$$
A_n (\Omega, \Omega_1) = \frac{\Omega - \Omega_1}{2} \left[2 arctg
\frac{\Omega}{\omega_0} - arctg \frac{\Omega}{W_n - \mu_n + \omega_0} - arctg
\frac{\Omega}{\mu_n + \omega_0} + \right.
$$
$$
\left. + arctg
\frac{\Omega_1}{\mu_n + \omega_0} - 2 arctg \frac{\Omega_1}{\omega_0} + arctg
\frac{\Omega_1}{W_n - \mu_n + \omega_0} \right],
$$
$$
B_n (\Omega, \Omega_1) =
- \frac{\mu_n + \omega_0}{2 [(\mu_n + \omega_0)^2 + \Omega_1^2]^2} \left[(\mu_n
+ \omega_0)^2 + 2 \Omega_1^2 - \Omega \Omega_1 \right] -
$$
$$ - \frac{W_n -
\mu_n + \omega_0}{2 [(W_n - \mu_n + \omega_0)^2 + \Omega_1^2]^2} \left[(W_n -
\mu_n + \omega_0)^2 + 2 \Omega_1^2 -\Omega \Omega_1 \right],
$$
$$
C_n (\Omega, \Omega_1) = \frac{1}{4} ln \frac{(W_n - \mu_n + \omega_0)^2 +
\Omega^2}{(\mu_n + \omega_0)^2 + \Omega^2} - \frac{1}{4} ln \frac{(W_n - \mu_n
+ \omega_0)^2 + \Omega_1^2}{(\mu_n + \omega_0)^2 + \Omega_1^2} +
$$
\begin{equation}
+ \Omega_1 (\Omega - \Omega_1) \frac{1}{2} \left[ \frac{1}{(\mu_n + \omega_0)^2
+ \Omega_1^2} - \frac{1}{(W_n - \mu_n + \omega_0)^2 + \Omega_1^2} \right].%27
\end{equation}

As follows from these formulas values of vertex functions depend significantly
on filling of the energy bands. At half-filling $\left( \mu _n = W_n/2 \right)$
all these expressions simplify significantly, the coefficient $C_n (\Omega,
\Omega_1)$, for instance, is equal to zero. \\

\begin{center}
{\bf 4. Temperature of superconducting transition $T_c$ and coefficient of
isotope effect $\alpha$.}
\end{center}

Assume in formulas (26), (27) $\Omega = 0$ and $\Omega_1 = \omega_0$
while having considered one - band non - adiabatic system \cite{Grimaldi},
\cite{Palistrant_2}. We introduce renormalization of electron - phonon
interaction couplings in accordance with their values
$$
\tilde \lambda_{11} = N_1 \tilde V_{11} (0, \omega_0) = \lambda_{11} +
\lambda_{11}^{2} \left[2P_{V_1} (Q_c, \omega_0) + P_{C_1} (Q_c,
\omega_0)\right],
$$
\begin{equation}
\lambda_{12} = V_{12} N_2 \qquad
\tilde \lambda_{nn}^{z} = \lambda_{nn} \left[1 + \lambda_{nn} P_{Vn} (Q_c,
\omega_0)\right]. %28
\end{equation}
As a result of these definitions the expression for mass operators $M_m
(\Omega)$ (21) and system of equations for $\Sigma_m (\Omega)$ has the following
form:
$$
M_m (\Omega) = - \tilde \lambda_{mm}^{z} \frac{1}{\beta N_m} \sum_{\vec p_1}
\sum_{\Omega_1} D (\Omega, \Omega_1) G_m (\vec p_1, \Omega_1) -
\frac{\lambda_{mn}}{\beta N_n} \sum_{\vec p_1} \sum_{\Omega_1} \times
$$
\begin{equation}
\times D(\Omega, \Omega_1)G_n (p_1, \Omega_1), (m \not = n) %29
\end{equation}
$$
\Sigma_1 (\Omega) = -\frac{\tilde \lambda_{11}}{\beta N_1} \sum_{\vec p_1}
\sum_{\Omega_1} D (\Omega, \Omega_1) F_1 (\vec p_1, \Omega_1) -
\frac{\lambda_{12}}{\beta N_2} \sum_{\vec p_1, \Omega_1} D (\Omega, \Omega_1)
F_2 (\vec p_1, \Omega_1),
$$
\begin{equation}
\Sigma_2 (\Omega) = -\frac{\tilde \lambda_{22}}{\beta N_2} \sum_{\vec p_1}
\sum_{\Omega_1} D (\Omega, \Omega_1) F_2 (\vec p_1, \Omega_1) - \frac{
\lambda_{21}}{\beta N_1} \sum_{\vec p_1} \sum_{\Omega_1} D (\Omega, \Omega_1)
F_1 (\vec p_1, \Omega_1). %30
\end{equation}
Introduce definitions $\Sigma_n (\Omega) = Z_n (\Omega) \Delta_n (\Omega)$.

Having substituted in (30) definitions of Green functions (11) in linear
approximation over $\Delta_n\,\,(T \sim T_c)$ and having performed integration
over energy in correspondence with (25) we obtain:
$$
\Delta_1 (\Omega) = \frac{\tilde \lambda_{11}}{Z_1} \frac{1}{\beta}
\sum_{\Omega_1} \frac{\omega_0^2}{(\Omega - \Omega_1)^2 + \omega_0^2}\,\,
\frac{\Delta_1(\Omega_1)}{|\Omega_1|} \varphi_1 (\Omega_1) +
$$
$$
+ \frac{\lambda_{12}}{Z_1} \frac{1}{\beta} \sum_{\Omega_1}
\frac{\omega_{0}^{2}}{(\Omega - \Omega_1)^2 + \omega_0^2}\,\,\frac{\Delta_2
(\Omega_1)}{|\Omega_1|} \varphi_2 (\Omega_1),
$$
$$
\Delta_2 (\Omega) = \frac{\tilde \lambda_{22}}{Z_2}\,\,\frac{1}{\beta}
\sum_{\Omega_1} \frac{\omega_0^2}{(\Omega - \Omega_1)^2 + \omega_0^2}\,\,
\frac{\Delta_2(\Omega_1)}{|\Omega_1|} \varphi_2 (\Omega_1) +
$$
\begin{equation}
+ \frac{\lambda_{21}}{Z_2 \beta}  \sum_{\Omega_1}
\frac{\omega_{0}^{2}}{(\Omega - \Omega_1)^2 + \omega_0^2}  \frac{\Delta_1
(\Omega_1)}{|\Omega_1|} \varphi_1 (\Omega_1),%31
\end{equation}
where
$$
\varphi_n (\Omega_1) = arctg \frac{W_n - \tilde \mu_n}{Z_n |\Omega_1|} +
arctg \frac{\tilde \mu_n}{Z_n |\Omega_1|},
$$
$$
Z_n = Z_n (0) = 1 - \stackrel{lim}{_{\Omega \rightarrow 0}} \frac{1}{\Omega}
Im M_n (\Omega) = 1 + \lambda_{nn}^{z} \frac{1}{2} \left[ \frac{W_n -
\mu_n}{W_n - \mu_n + \omega_0} + \frac{\mu_n}{\mu_n + \omega_0} \right]+
$$
\begin{equation}
+ \lambda_{nm} \frac{1}{2} \left[\frac{W_m - \mu_m}{W_m - \mu_m + \omega_0} +
\frac{\mu_m}{\mu_m + \omega_0}\right], \quad \tilde \mu_n = \mu_n - Re M_n
(0),\quad(m \not = n).%32
\end{equation}
While studying the system of equation (31) we utilize approximation which was
used in theory of superconductivity with taking into account the lagging effect
\cite{Mc Millan}
\begin{equation}
\frac{\omega_0^2}{(\Omega - \Omega_1)^2 + \omega_0^2} \longrightarrow
\frac{\omega_0^2}{\Omega^2 + \omega_0^2} \frac{\omega_0^2}{\Omega_1^2 +
\omega_0^2} %33
\end{equation}
and introduce the following notation:
\begin{equation}
\Delta_n (\Omega) = \frac{\omega_0^2}{\Omega^2 + \omega_0^2} \Delta_n ; \quad
\Phi_n (T_c) = \frac{1}{\beta_c} \sum_{\Omega_1} \frac{\omega_0^4}{(\Omega_1^2 +
\omega_0^2)^2} \frac{1}{|\Omega_1|} \varphi_n (\Omega_1),%34
\end{equation}
Having substituting (33), (34) into the system of equations (31), we represent
this system of equations to the form:
\begin{equation}
\Delta_1 = \frac{\tilde \lambda_{11}}{Z_1} \Delta_1 \Phi_1 (T_c) +
\frac{\lambda_{12}}{Z_1} \Delta_2 \Phi_2 (T_c),\quad \Delta_2 =
\frac{\lambda_{21}}{Z_2} \Delta_1 \Phi_1 (T_c) + \frac{\tilde
\lambda_{22}}{Z_2} \Delta_2 \Phi_2 (T_c). %35
\end{equation}
Temperature of superconducting transition $T_c$ is determined from the condition
of consistency of the system of equations (35). Having supposed the determinant
of this system to be equal to zero, we derive
\begin{equation}
(\bar \lambda_{11} \bar \lambda_{22} - \bar \lambda_{12} \bar \lambda_{21})
\Phi_1 (T_c) \Phi_2 (T_c)  - \bar \lambda_{11} \Phi_1 (T_c) - \bar \lambda_{22}
\Phi_2 (T_c) + 1 = 0,%36
\end{equation}
were
$\bar \lambda_{11} = \tilde \lambda_{11}/Z_1 ;\quad  \bar \lambda_{22} =
\tilde \lambda_{22}/Z_2;\quad \bar \lambda_{12} = \lambda_{12}/Z_1;
\quad \bar \lambda_{21} = \lambda_{21}/Z_2$

The function $\Phi_n(T_c)$ can be easily represented in the form:

$$
\Phi_n (T_c) = \pi T_c \sum_{\Omega_1} \frac{\omega_0^2}{\Omega_1^2 + \omega_0^2}
\frac{1}{|\Omega_1|} - T_c \sum_{\Omega_1} \frac{\omega_0^2}{\Omega_1^2 +
\omega_0^2} \frac{1}{|\Omega_1|} \left[ arctg \frac{|\Omega_1|}{\bar W_n -
\bar \mu_n} + \right.
$$
\begin{equation}
\left. + arctg \frac{|\Omega_1|}{\bar \mu_n} \right] - T_c \sum_{\Omega_1}
\frac{\omega_0^2 |\Omega_1|}{(\Omega_1^2 + \omega_0^2)^2} \varphi_n (\Omega_1).
%37
\end{equation}
First term of this expression contains logarithmic singularity over the quantity
$T_c$ and in the weak coupling approximation $(T_c/\omega_0 \ll 1)$ gives
\begin{equation}
\pi T_c \sum_{\Omega_1} \frac{\omega_0^2}{\Omega_1^2 + \omega_0^2}
\frac{1}{|\Omega_1|} \approx ln \frac{2 \omega_0 \gamma}{\pi T_c}.%38
\end{equation}
In the same approximation we can perform integration over frequency in infinite
limits (25) as in the case $T = 0$ in all rest terms. In this way for the
function $\Phi_n (T_c)$ we obtain:
\begin{equation}
\Phi_n (T_c)  = \xi_c + f_n,\quad \xi_c = ln \frac{2 \omega_0 \gamma}{\pi
T_c},%39
\end{equation}
where
$$
f_n = - \frac{1}{2} \left[ln(1 + m_n) + ln (1 + m_{n}^{'}) - \frac{1}{2} +
\frac{1}{4} \left(\frac{m_n}{1 + m_n} + \frac{m_{n}^{'}}{1 +
m_{n}^{'}}\right)\right],
$$
\begin{equation}
m_n = \frac{\omega_0}{\bar W_n - \bar \mu_n},\quad m_{n}^{'} =
\frac{\omega_0}{\bar \mu_n},\quad \bar W_n = \frac{W_n}{Z_n},\quad \bar \mu_n =
\frac{\mu_n}{Z_n}. %40
\end{equation}
Having substituted (39) into the equation (36) we lead the latter one to the
form:
\begin{equation}
\bar a \xi_c^2 - \bar b \xi_c + \bar c = 0,%41
\end{equation}
where
$$
\bar a = \bar \lambda_{11} \bar \lambda_{22} - \bar \lambda_{12} \bar
\lambda_{21};\quad \bar b = \bar \lambda_{11} + \bar \lambda_{22} - \bar a
(f_1 + f_2),
$$
\begin{equation}
\bar c = 1 - \bar \lambda_{11} f_1 - \bar \lambda_{22} f_2 + \bar a f_1
f_2.%42
\end{equation}
Having based on the (39) and (41) we obtain
\begin{equation}
T_c = \frac{2 \omega_0 \gamma}{\pi} e^{- \xi_c},\quad \xi_c = \frac{\bar b
\pm \sqrt{\bar b^2 - 4 \bar a \bar c}}{2 \bar a}.%43
\end{equation}
This expression coincides formally with the case of usual two-band
superconductors \cite{Moskalenko}, \cite{Moskalenko_2}. But this expression differs
significantly because the quantities which enter in it are renormalized by
additional contributions that are determined by non-adiabaticity of system. In
order to go to one-band case it is sufficient to put $N_2 = 0$. Having following
this way we derive the temperature of superconducting transition $T_{c_0} =
T_c|_{N_2 = 0}$ \cite{Palistrant_2}.
\begin{equation}
T_{c_0} = \frac{2 \omega_0 \gamma}{\pi \sqrt{e}} \frac{[(\bar W_1 - \bar
\mu_1) \bar \mu_1]^{1/2}}{[(\bar W_1 - \bar \mu_1 + \omega_0) (\bar \mu_1 +
\omega_0)]^{1/2}} exp \biggl\{- \frac{1}{\bar \lambda_{11}} + \frac{1}{4}
\left(\frac{1}{\bar W_1 - \bar \mu_1 + \omega_0} + \frac{1}{\bar \mu_1 +
\omega_0}\right) \biggr\}.%44
\end{equation}
At half-filling of energy band $\mu_1 = W_1/2$ this expression goes into one
which was obtained in article \cite{Grimaldi}:
\begin{equation}
T_{c_0}^{0} = \frac{2 \omega_0 \gamma}{\pi \sqrt{e} (1 + m)}  exp \biggl\{
- \frac{1}{\bar \lambda_{11}} + \frac{1}{2} \frac{m}{m + 1} \biggr\},%45
\end{equation}
where $m = \frac{\textstyle {2 \omega_0}}{\textstyle{W_1}}$\\
Therefore, the formula (43) shows the influence of overlapping of energy bands
on Fermi surface and non-adiabaticity $(\varepsilon_F \sim \omega_0)$ on the
temperature of superconducting transition at arbitrary filling of energy bands.

Having based on the formula(43) the expression for the coefficient of isotope
effect $\alpha$ can be represented in the form:
\begin{equation}
\alpha = - \frac{d ln T_c}{d ln M} = \frac{1}{2} \left[1 + \frac{d ln
(T_c/\omega_0)}{d ln \omega_0}\right] = \frac{1}{2} \left[1 -
\frac{d \xi c}{d ln \omega_0}\right],%46
\end{equation}
where
$$
\frac{d \xi_c}{d ln \omega_0} = - \frac{\bar b \pm \sqrt{\bar b^2 - 4 \bar a
\bar c}}{\bar a^2} \frac{d \bar a}{d ln \omega_0} + \frac{1}{2 \bar a} \times
$$
\begin{equation}
\times \biggl\{\frac{d \bar b}{d ln \omega_0} \pm \frac{1}{\sqrt{\bar b^2 - 4
\bar a \bar c}} \left[\frac{d \bar b}{d ln \omega_0} - 2 \bar a \frac{d \bar
c}{d ln \omega_0} - 2 \bar c \frac{d \bar a}{d ln \omega_0}\right]\biggr\}%47
\end{equation}

As follows from (46) and (47) the coefficient $\alpha$ differs from $1/2$, which
characterizes the ordinary superconductors, through dependence of the quantities
$\bar a,\,\bar b,\,\bar c$ on $\omega_0$. This dependence is caused by non-
adiabaticity of system and by delay in the system with electron-phonon
interaction.

Note that the sign in formulas (43) and (47) is chosen from the condition
$\xi_c > 0$ and maximum value of the quantity $T_c$ as it has been done in
the case of adiabatic systems.\\
\\
\begin{center}
{\bf 5. Numerical calculations and discussions.}\\
\end{center}

Theory of superconductivity for two-band non-adiabatic system by exceeding the
bounds of Migdal theorem \cite{Migdal} in linear over non-adiabaticity
approximation is built in this capter. Non-adiabaticity here means that having
determined the mass operators, diagonal $M_n$ and non-diagonal $\Sigma_n$ in
electron Green function of two-band system $(n = 1, 2)$ the additional in
comparison with Eliashberg-Migdal theory diagrams with intersection of two
lines of electron-phonon interaction, which correspond to vertex functions
$P_{V_n}$ as well as to "intersecting" ones $P_{C_n} (n = 1, 2)$, are taken into
consideration. Applied weak coupling approximation $(T_C/\omega_0 ;
\Delta_n/\omega_0 \ll 1)$ allows us to obtain analytical formulas for these
functions and show that their behavior is determined by the value of transferred
momentum $q$ of elctron-phonon interaction. At low values of cut-off momentum
of electron-phonon interaction $q_c (q_c \ll 2_{p_F})$ functions $P_{V_n}$ and
$P_{C_n}$ give positive contribution and at $q \sim 2_{p_F}$ they give negative
contribution. Lowness of the quantity $q_c$ is provided, for instance, by
presence of strong electron correlations in the system \cite{Kulic},
\cite{Zeyher}, or by quasi-one- dimensional dispersion law of electron's energy
\cite{Palistrant_1}.

For numerical calculations a simplified case $m = m_1 = m_2 (m_n =
2\,\,\omega_0/W_n)$ and half filling of energy bands $\mu_n = W_n/2$ were
considered.

The theory parameters are: $\lambda_{11} = 0.5$; $\lambda_{22} = 0.3$;
$\lambda_{12} = 0.1$; $\lambda_{21} = 0.05$.

According to the definitions (36), (28) and (32) these parameters
are renormalized due to the non - adiabatic effects and strong electron
correlations. The values of renormalized $\lambda_{nm}$ at different values
of Migdal parameter $m$ are presented in the Table 1 along with the values of
the functions $P_{V_1}$ and $P_{C_1}$.
\begin{center}
{\bf Table 1.}
\end{center}
\begin{center}
\begin{tabular}{|r|r|r|r|r|r|r|r|}\hline
& & & & & & & \\
$Q_c = 0,1$ & $\bar m$ & $\quad P_{V_1}\quad$ & $\quad\,\, P_{C_1}\,\,$ & $\quad
\bar \lambda_{11}$ & $\quad \bar \lambda_{22}$ & $\quad \bar \lambda_{12}$ &
$\quad \bar \lambda_{21}$ \\ \hline \end{tabular}\\
\end{center}
\begin{center}
\begin{tabular}{rrrrrrrr}
\qquad \quad\,\, $\,\,$ & $0$ & $0$ & $0$ & $\,\,0,312$ & $\,\,\,0,24$ &
$0,065$ & $0,038$\\ $\,\,$ & $0,1$ & $\,\,0,679$ & $\,\,\,0,656$ & $0,59$ &
$0,35$ & $0,058$ & $0,036$\\ $\,\,$ & $0,3$ & $\,\,0,565$ & $\,\,\,0,553$ &
$0,584$ & $0,344$ & $0,064$ & $0,038$\\ $\,\,$ & $0,6$ & $\,\,0,426$ &
$\,\,\,0,425$ & $0,568$ & $0,333$ & $0,069$ & $0,04$\\ $\,\,$ & $0,9$ &
$\,\,0,342$ & $\,\,\,0,342$ & $0,556$ & $0,32$ & $0,073$ & $0,042$\\
\end{tabular}
\end{center}
\begin{center}
\begin{tabular}{|r|r|r|r|r|r|r|r|}\hline
& & & & & & & \\
$Q_c = 0,9$ & $\bar m$ & $\quad P_{V_1}\quad $ & $\quad\,\, P_{C_1\,\,}$ &
$\quad \bar \lambda_{11}$ & $\quad \bar \lambda_{22}$ & $\quad \bar
\lambda_{12}$ & $\quad \bar \lambda_{21}$ \\ \hline
\end{tabular}\\
\end{center}
\begin{center}
\begin{tabular}{rrrrrrrr}
\qquad \quad\,\,$\,\,$ & $0$ & $0$ & $0$ & $0,312$ & $0,24$ & $0,065$ &
$0,038$\\ $\,\,$ & $0,1$ & $-0,053$ & $-0,044$ & $0,315$ & $0,223$ & $0,065$ &
$0,038$\\ $\,\,$ & $0,3$ & $-0,012$ & $-0,096$ & $0,322$ & $0,227$ & $0,068$ &
$0,039$\\ $\,\,$ & $0,6$ & $-0,008$ & $-0,112$ & $0,339$ & $0,236$ & $0,073$ &
$0,041$\\ $\,\,$ & $0,9$ & $0,005$ & $-0,113$ & $0,36$ & $0,245$ & $0,076$ &
$0,042$\\
\end{tabular}
\end{center}

As it follows from the Table 1 the behavior of renormalized electron - phonon
interaction constants is not a single - valued one and is determined in the
same time both the value of cut - off momentum of electron - phonon interaction
$Q_c$ and the value of Migdal parameter $m$. The ration $T_c/\omega_0$ as a
function of $m$ is presented on the Fig. 1. Here we have considered the
definition (43) and the obtained results multiplied by $\sqrt{\varepsilon}$.
Such correction is necessary because the approximation (33) decrease the $T_c$
with the coefficient $e^{-1/2}$ \cite{Combescot} . The curve 2 on this figure
correspond to the case of two - band  adiabatic system, curves $1$ and $1'$
correspond to the case of non adiabatic two - band system at $Q_c = 0,1$ and
$Q_c = 0,9$ respectively. The comparison of these curves shows that the non -
adiabatic effects and strong electron correlations considerably increase the
temperature of superconducting transition $T_c$ at $Q_c \ll 1$. Moreover, the
biggest value of $T_c$ is reached in the area of small $m \, (m \sim 0.1)$. At
$Q_c \sim 1$ the $T_c$ decreases and weakly depends on parameter $m$.

These results point on the fact that in the systems with small values of the
ratio Debye energy on Fermi energy the contribution of top approximations to
the superconduc-\\
tivity can be essential if the trassmitted momentum is small.

The substances with small values of $m\, (m \sim 0.1)$ particularly are two -
band superconductors $MgB_2$. The strong electron correlations presence or
the peculiarities in electron enrgy spectrum in such systems can essentially
strength the influence of non adiabatic effects on superconductivity.

The dependency of ratio $T_c/\omega_D$ on coefficient $\delta$ which determines
the proportional increasing of all electron - phonon interaction constants in
two - band model $(\lambda_{nm}^{'} = \delta \lambda_{mn})$ is presented on
Fig.2. The cases $m = 0,2;\,\,Q_c = 0,1$ (curve 1) and $m = 0,2;\,\, Q_c = 0,9$
(curve 2) are considered. Increase of $T_c$ followes from this figure with
increasing the coefficient $\delta$ in both cases.

The dependency of isotopic coefficient $\alpha$ on Migdal parameter $m$ is
shown on Fig. 3. At $Q_c \ll 1$ and $m < 0.1$ the coefficient $\alpha > 0.5$,
then this quantity  essentially decreases with the increasing of $m$ and can
achieve the values $\sim 0.2$ (curve 1). At $Q_c \sim 1$ the value of $\alpha$
becomes $\approx 0.5$ (weakly decreases then increases with the increasing of
$m$).

In this way, the non-adiabatic effects in two - band superconductors at small
values of cut - off momentum of electron - phonon interaction $Q_c$ allow to
attain high values of superconducting transition temperature $T_c$ and small
values of isotopic effect. Such picture is observed in high temperature oxide
metals.

The presented above numerical results are performed for the $\lambda_{11} >
\lambda_{22} > \lambda_{12} > \lambda_{21}$. These results correspond
qualitatively to the case of one - band superconductor \cite{Grimaldi}.

In the limit case $\lambda_{11} = \lambda_{22} = 0$,\,$\lambda_{12}$ and
$\lambda_{21} \not = 0$ the examined non adiabatic effects do not influency
the superconductivity, because in this case the interband interaction constants
$\lambda_{nm}$ are not renormalized due to vertex functions consideration.

We have to emphasize that the possibility of appearance of high temperature
supercon-\\
ductivity in the systems with strong electron correlations on the base
of phonon supercon-\\
ducting mechanism was discussed in a chain of articles (see, for instance
\cite{Kim_1} - \cite{Moskalenko_3}).


\begin{thebibliography}{99}
\bibitem{Berdnorz} J. G. Berdnorz and K. A. Miiller, Z. Phys. B, {\bf 64}, 189
(1986).%1

\bibitem{Moskalenko} V. A. Moskalenko,  {\it Fiz. Met.Metalloved};
{\bf 8}, 503 (1959); {\it Phys. Met. and Metallog.} {\bf 8}, 25 (1959) %2

\bibitem{Suhl} H. Suhl, B. T. Matthias, and L. R. Walker,
{\it Phys. Rev. Lett.} {\bf 3}, 552 (1959). %3

\bibitem{Palistrant} M. E. Palistrant, arXiv: cond - mat / 0305491 (2003);
Moldavian Journal of the Physical Sciences, 3 N2 (2004) %4

\bibitem{Kon} L. Z. Kon, arXiv: cond - mat /0309707 (2003).%5

\bibitem{Moskalenko_1} V. A. Moskalenko, M. E. Palistrant and V. M. Vakalyuk,
{\it Usp. Fiz. Nauk}, {\bf 161}, 155 (1991); {\it Sov. Phys. Usp.}
{\bf 34}, 717 (1991); arXiv: cond - mat/03099671. %6

\bibitem{Migdal} A. B. Migdal, {\it Zh. Eksp. Teor. Fiz.},
{\bf 34},1438 (1958). %7

\bibitem{Danilenko} O. V. Danilenko and O. V. Dolgov, {\it Preprint cond - mat
0007189}.%8

\bibitem{Kulic} M. L. Kulic and R. Zeyher, {\it Phys. Rev. B}, {\bf 49},
4395 (1994).%9

\bibitem{Zeyher} R. Zeyher and M. L. Kulic , {\it Phys. Rev. B}, {\bf 53},
2850 (1996).%10

\bibitem{Pietronero} L. Pietronero, S. Strassler and S. Grimaldi, {\it Phys.
Rev.  B}, {\bf 52}, 10516 (1995).%11

\bibitem{Grimaldi} S. Grimaldi, L. Pietronero and Strassler, {\it Phys.
Rev.  B}, {\bf 52}, 10530 (1995).%12

\bibitem{Palistrant_1} M. E. Palistrant, {\it Fiz. Nizk. Temp.}, 500  (2005);
{\it Low Temperature Physics}, {\bf 31}, 738 (2005).%13

\bibitem{Abrikosov} A. A. Abrikosov, Y. C. Campuzano and K. Gofron, {\it
Physica C}, {\bf 214}, 73 (1993).%14

\bibitem{Palistrant_2} M. E. Palistrant and F. G. Kochorbe, {\it Fiz. Nizk.
 Temp.}, {\bf 26} 557 (2000); {\it J. of Superconductivity: Incoparating
 Novel Magnetizm}, {\bf 15}, 113 (2002); {\it J. Phys. Condens Matter},
{\bf 15}, 3267 (2003); {\it Low Temp. Phys.}, {\bf 29}, 1173 (2003); {\it
International Journal of Modern Physics B}, {\bf 17}, 2545 (2003).  %15

\bibitem{Palistrant_3} M. E. Palistrant, {\it Fiz. Nizk.
Temp.}, {\bf 28} 157 (2002); {\it Teoret. i Mat. Fizika}, {\bf 135}, 137
(2003); {\it Theoret. Mathem. Phys.}, {\bf 135}, 566 (2003).%16

\bibitem{Nagamatsu} J. Nagamatsu, N. Nakagawa, T. Muranaka, Y. Zenitani and J.
Akimistu, Nature (London), {\bf 463}, 401 (2001).%17

\bibitem{Confield} P. C. Confield, S. L. Bud'ko and D. K. Finemore,
{\it Physica C}, {\bf 385}, 1 (2003).%18

\bibitem{Bouquet} F. Bouquet, J. Wang, I. Sheikin et al.,{\it Physica C},
{\bf 385}, 192 (2003).%19

\bibitem{Mishonov} T. Mishonov, S. L. Drechsler and E. Penev, {\it Modern
Physics Letters B}, {\bf 17}, 755 (2003); arXiv: cond - mat/0209192 .%20

\bibitem{Mishonov_1} T. Mishonov, E. Penev, J. O. Indekeu and V. I. Pokrovsky,
arXiv: cond - mat/0209342 ; {\it Phys. Rev. B}, {\bf 68} 104517 (2003).%21

\bibitem{Dolgov} O. V. Dolgov, R. K. Kremer at al., {\it Phys. Rev. B}, {\bf
72} 024504 (2005).%22

\bibitem{Aleksandrov} A. S. Aleksandrov, Preprint cond - mat/0102189 .%23

\bibitem{Krakauer} H. Krakauer and E. Pickett, {\it Z. Phys.}, {\bf
60}, 1665 (1988).%24

\bibitem{Herman}J. F. Herman, R. V. Kasowski and W. G. Hsu, {\it Phys. Rev. B},
{\bf 36}, 6904 (1987).%25

\bibitem{Kortus} J. Kortus, I. I. Mazin, K. D. Belashchenko, V. P. Antropov
and L. L. Boyer, {\it Phys. Rev. Lett.}, {\bf 86} 4656 (2001).%26

\bibitem{An} J. M. An and W. E. Pickett, {\it Phys. Rev. Lett.}
{\bf 86} 4366 (2001).%27

\bibitem{Abrikosov_1} A. A. Abricosov, L. P. Gor'kov and I. E. Dzyaloshinskii,
{\it Quantum Field Theoretical Methods in Statistical Physics} Moscow, Nauka
(1962); {\it Prentice - Hall, Englewood Cliffs}, N7 (1963).%28

\bibitem{Kochorbe} F. G. Kochorbe and M. E. Palistrant,{\it Zh. Eksp. Teor.
Fiz.}, {\bf 104}, 3084 (1993); [{\it JETP} {\bf 77}, 442 (1993)]; {\it Teoret.
Mathem. Phis.} {\bf 96}, 459 (1993); [{\it Theoret. Mathem. Phys.} {\bf
96}, 1083 (1993)].%29

\bibitem{Palistrant_4}  M. E. Palistrant,{\it Intern. Joyrnal of Modern
Physics B} {\bf 19}, 929 - 970 (2005).%30

\bibitem{Mc Millan} W. L. Mc Millan, {\it Phys. Rev.} {\bf 167}, 331 (1968).%31

\bibitem{Moskalenko_2} V. A. Moskalenko, L. Z. Kon and M. E. Palistrant,
{\it Low - Temperature Properties of Metals with Band - Spectrum
singularities} [in Russian], Shtiintsa, Kishinev (1989).%32

\bibitem{Palistrant_5}  M. E. Palistrant and F. G. Kochorbe ,{\it Physica C},
351 (1992).%33

\bibitem{Combescot} R. Combescot,{\it Phys. Rev. B} {\bf 42},
7810 (1990).%34

\bibitem{Kim_1} J. H. Kim and Tesanovic,{\it Phys. Rev. Lett.} {\bf 71}, 4218
(1993).%35

\bibitem{Lanzara} A. Lanzara, P. V. Bogdanov, X. I.Zhow at al. {\it Nature}
{\bf 412}, 510 (2001).%36

\bibitem{Moskalenko_3} V. A. Moskalenko, P. Entil, M. Marinaro, D. F. Digor,
{\it Zh. Eksp. Teor. Fiz.} {\bf 124}, 700 (2003) [{\it JETP} {\bf 97},
632 (2003)].%37






\end{thebibliography}
\end{document}